\documentclass[aps,prb,twocolumn,superscriptaddress,showpacs,amsmath,amssymb,
amsfonts,balancelastpage]{revtex4-2}
\usepackage{amsmath}
\usepackage{amssymb}
\usepackage{amsfonts}
\usepackage[justification=raggedright]{caption}
\usepackage{subcaption}
\usepackage{graphicx}
\usepackage{epsfig}
\usepackage{hyperref}
\usepackage{epstopdf}
\usepackage{gensymb}
\usepackage{color }
\usepackage{color, soul }
\usepackage{xcolor}
\hypersetup{
	colorlinks=true,
	urlcolor=blue,
	citecolor=blue,
	linkcolor=blue,}
\begin{document}
		\title{Observation of Weak Anti-localization in thin films of the Topological Semimetal Candidate PdSb$_{2}$}
	
	\author{Shama}
	\affiliation{Department of Physical Sciences, Indian Institute of Science Education and Research, Knowledge city, Sector 81, SAS Nagar, Manauli PO 140306, Mohali, Punjab, India}

	\author{Aastha Vasdev}
	\affiliation{Department of Physical Sciences, Indian Institute of Science Education and Research, Knowledge city, Sector 81, SAS Nagar, Manauli PO 140306, Mohali, Punjab, India}

	\author{Dinesh Kumar}
	\affiliation{Central Research Facility, Indian Institute of Technology Delhi, Hauz Khas, New Delhi - 110016, India}

\author{Goutam Sheet}
	\affiliation{Department of Physical Sciences, Indian Institute of Science Education and Research, Knowledge city, Sector 81, SAS Nagar, Manauli PO 140306, Mohali, Punjab, India}
	
	\author{Yogesh Singh} \email{yogesh@iisermohali.ac.in}
	\affiliation{Department of Physical Sciences, Indian Institute of Science Education and Research, Knowledge city, Sector 81, SAS Nagar, Manauli PO 140306, Mohali, Punjab, India}
	
\begin{abstract}
 We report results of a magneto-transport study on thin films of the topological semi-metal candidate PdSb$_{2}$ (PS).  We observe a positive correction to magneto-conductivity at low temperatures, which is a signature of weak anti-localization (WAL). We analyze the WAL data within the Hikami-Larkin-Nagaoka (HLN) theory and extract the dephasing length (L$_{\phi}$) whose temperature dependence reveals the various phase relaxation mechanisms. From the WAL effect, we also extract $\alpha$ (the number of transport channels).  The evolution of $\alpha$ with temperature and film thickness reflects how the coupling between different conducting channels (Topological surface channels and bulk channels) changes. Additionally, the electron-electron interaction (EEI) effect was observed in the temperature-dependent conductivity at low temperatures. From the EEI effect, we get an alternate estimate of the number of transport channels and obtain a value similar to that obtained from the analysis of the WAL effect.  This suggests that in disordered films, the EEI effect can be used to get information about the coupling of the topological surface states with each other and with bulk states.
\end{abstract}
		\maketitle
	
\section{Introduction}
The discovery of topological materials such as topological insulators, topological semimetals, etc., has opened up new avenues of research in condensed matter physics.\cite{Hasan,Moore,Burkov,Wan,Armitage} Topological insulators (TIs) are new states of matter which have gapped insulating bulk states with gapless conductive surface states protected by time-reversal symmetry (TRS).\cite{Hasan,Moore} In Dirac Semimetals (DS), the crossing of conduction and valence band occurs at discrete points in the Brillouin Zone, which is protected by time-reversal symmetry (TRS) and inversion symmetry (IS).\cite{Armitage,Young,Wang,Zhou} This crossing leads to four-fold degeneracy at discrete points. If either TRS or IS, gets broken, the Dirac point with four-fold degeneracy splits into two Weyl points with two-fold degeneracy, forming a Weyl semimetal (WS).\cite{Armitage,Lv}  The non-trivial nature of topological materials result in exotic transport properties such as extremely large magneto-resistance, chiral anomaly induced negative MR, high carrier mobility, and anomalous Hall effect.\cite{Zhang,Xiong,Shekhar,Liang} In high-energy physics, fermions are constrained by Poincaré invariance. However, in condensed-matter systems, fermions in crystals are restricted by the symmetries of the 230 crystal space groups rather than by Poincaré invariance, giving rise to the possibility of realizing other types of fermions which have no counterparts in high-energy physics.\cite{Tang,Jiang,Chang,Bradlyn}
	
Recently, based on density functional theory (DFT) calculations, Bradlyn et al. have predicted the existence of such unconventional fermions having three, six, and eightfold degenerate band crossing in various materials.\cite{Bradlyn} The six and eight-fold band crossings are stabilized by non-symmorphic symmetry, while the three-fold band crossing is stabilized by both non- symmorphic and symmorphic symmetry. These higher-fold band crossing has been observed in many materials such as MoP, CoSi, AlPt etc.\cite{B,Sanchez,YChen} PdSb$_{2}$ is one such material which has been proposed to consist of fermions with six-fold degenerate band crossing. Previous studies have confirmed the existence of sixfold degenerate band crossings by angle-resolved photoemission spectroscopy (ARPES) and density functional theory (DFT). \cite{Sun,Nitesh,Tyler} Also, R. Chapai.\textit{et al.} have probed the presence of massless particles with non-trivial Berry phase by de Hass-van oscillations in PdSb$_{2}$ single crystals.\cite{Chapai} Therefore, it is possible to create three or two-fold Weyl nodes in PdSb$_{2}$ by breaking the time-reversal symmetry either by applying an external magnetic field or doping.\cite{Vergniory}
	
The Berry phase associated with surface electrons in topological materials leads to the suppression of backscattering and hence the immunity towards weak localization.\cite{Roushan,Ostrovsky,Lu,He} This gives rise to a quantum correction to conductivity whose temperature and magnetic field dependences are similar to the weak anti-localization (WAL) effect in 2D systems.\cite{McCann,Tkachov,Lee,Bergmann} However, the WAl effect is suppressed by the application of an external magnetic field, which destroys the time-reversal symmetry resulting in localization of the electron. WAL has been described by Hikami, Larkin, and Nagaoka (HLN) in two-dimensional systems.\cite{Hikami,Fang} Another quantum phenomena observed in many Topological materials at low temperatures is the electron-electron interaction effect (EEI).\cite{JWang,Liu,Chen,Takagaki,Chiu,Roy}  From the EEI effect, one can get information about the number of conducting transport channels.\cite{JWang} Transport experiments often involve non-negligible contribution from the bulk states, which complicates probing surface states electronically. Therefore, the fabrication of thin films is an effective technique to reduce the bulk contribution by increasing the surface-to-volume ratio.  Additionally, it has been shown recently that disorder in thin films can lead to a much reduced mobility of bulk carriers, which in turn enhances the relative surface contribution. \cite{Shama}

 Here, we report the growth of PdSb$_{2}$ (PS) thin films of two thicknesses using the pulsed laser deposition technique (PLD) and a study of their magneto-transport properties. We have observed a sharp cusp around B = 0 under applied transverse magnetic field, revealing the presence of weak anti-localization (WAL).  By applying the HLN theory, we have extracted the dephasing length (L$_{\phi}$) and found a deviation from the behaviour expected from just the Nyquist electron-electron dephasing mechanism. This indicates the presence of some other phase relaxation mechanism such as electron-phonon scattering.\cite{Wu,Bird,Sergeev,Reizer} We also estimated the number of conducting transport channels from both the WAL effect and the EEI effect.  We are thus able to track the evolution of the coupling between various conduction channels (Topological surface states and bulk states) with film thickness and temperature.

\section{Methods}
\begin{figure}
	\centering
	\includegraphics[width= 1\linewidth, height=0.3\textheight]{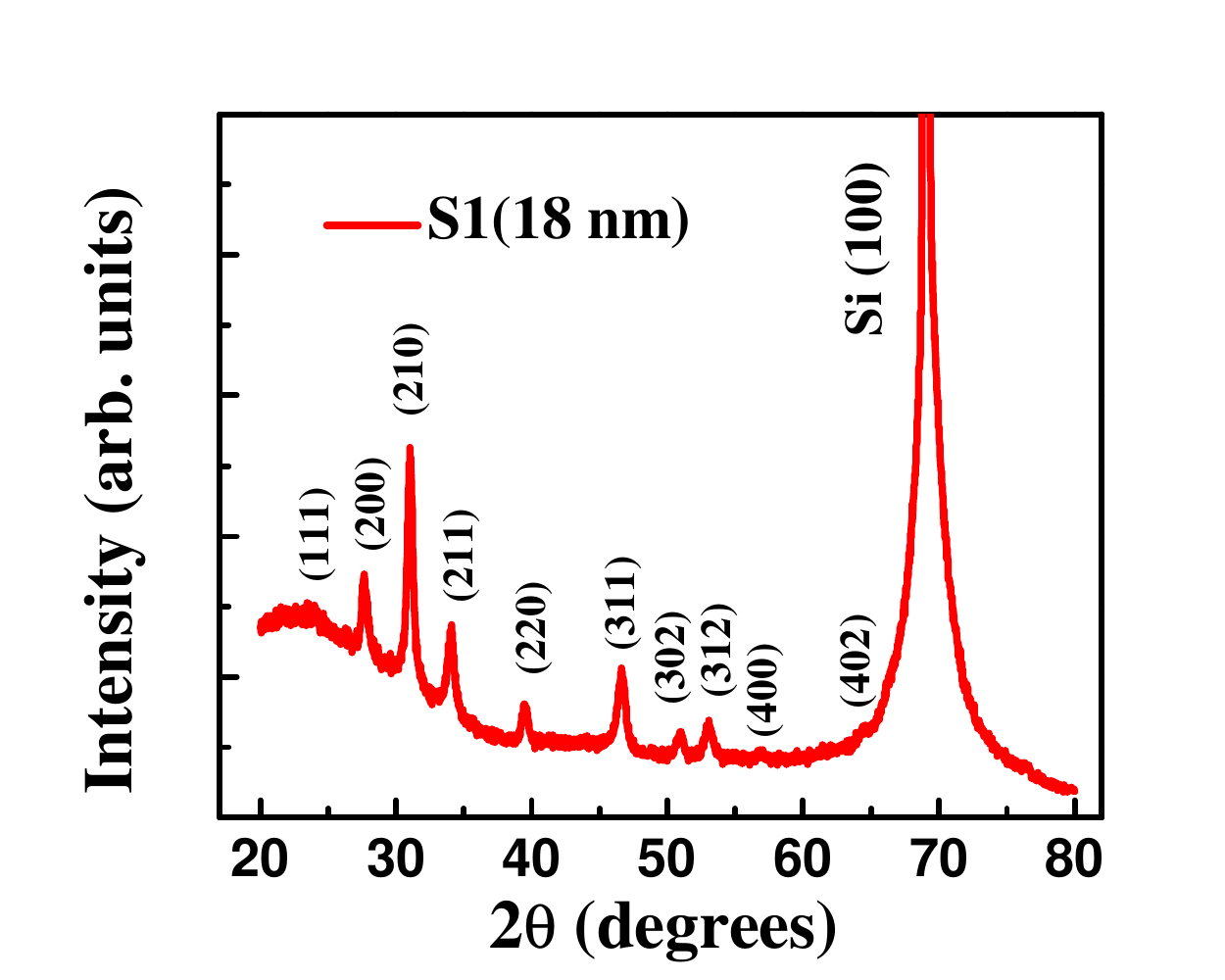}
	\caption{X-ray diffraction data for S1 thin films of PdSb$_{2}$}
	\label{fig:xrd-data}
\end{figure}

\begin{figure}
	\centering
	\includegraphics[width=1.1\linewidth, height=0.35\textheight]{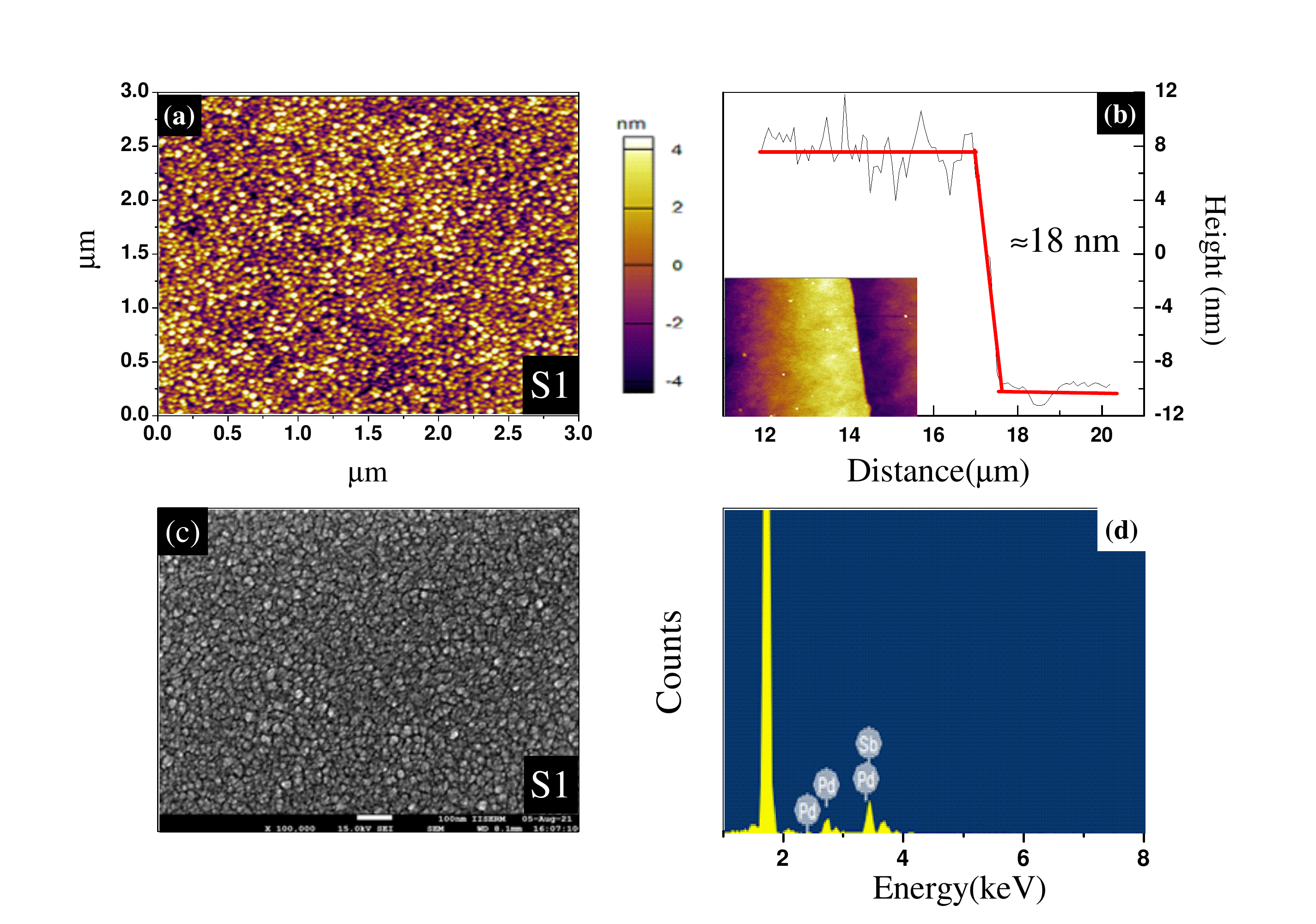}
	\caption{(a) Atomic force microscope topographic image of the surface of S1 thin films of PdSb$_2$. (b) Height profile of S1 giving a thickness $\approx$ 18 nm. (c) Scanning electron microscope images of S1 films. (d) Energy dispersive spectroscopy data giving the chemical composition of S1.}
	\label{fig:afm-sem}
\end{figure}

 Thin films of PS with different thickness were grown on a (100) oriented silicon substrate using a pulsed laser ( KrF excimer, $\lambda$ = 248 nm) deposition technique in Argon atmosphere. The laser ablation was performed on a polycrystalline PS target formed by solid-state synthesis. The thin film growth of PS was carried out at a pressure of $\approx 10^{-1}$~mbar and substrate temperature 120$^{\circ}$C.
X-ray diffraction on a Bruker D8 Advance diffractometer system with Cu-K$\alpha$ radiation was used to determine the phase purity of PS thin films. The stoichiometry of PS thin films was confirmed using energy dispersive spectroscopy using a scanning electron microscope. The film topography and thickness was measured using an atomic force microscope (AFM).   The PS thin films with thickness 18 nm and 10 nm are designated as S1, S2 respectively. The longitudinal and Hall resistances were measured in a Physical Property Measurement System (PPMS-Cryogenics ) equipped with a 14T magnet.

\section{Results and Discussion}
Figure~\ref{fig:xrd-data} shows the x-ray diffraction (XRD) pattern of a thin film with a thickness of $\approx$ 18 nm. The XRD pattern confirms the cubic crystal structure with space group P 2$_{1}$/a$\overline{3}$ similar to bulk samples reported previously.\cite{Sun,Nitesh,Tyler,Chapai} Figure~\ref{fig:afm-sem}(a) shows the atomic force microscope (AFM)  image of S1 thin film, which indicates the granular growth of the film. Figure~\ref{fig:afm-sem}(b) shows the height profile of S1 thin film with thickness $\approx$ 18nm. The inset shows the topography image. Figure~\ref{fig:afm-sem}(c) shows the scanning electron microscopy (SEM) image of the thin film S1. Figure~\ref{fig:afm-sem}(d) shows the energy-dispersive
x-ray spectroscopy (EDS) data where Pd and Sb peaks are observed. The atomic ratio of Pd and Sb confirms the stoichiometry of the compound. From characterization data, it can be concluded that we have successfully grown the polycrystalline films of PdSb$_{2}$.

\begin{figure}
	\centering
	\includegraphics[width=1.1\linewidth, height=0.22\textheight]{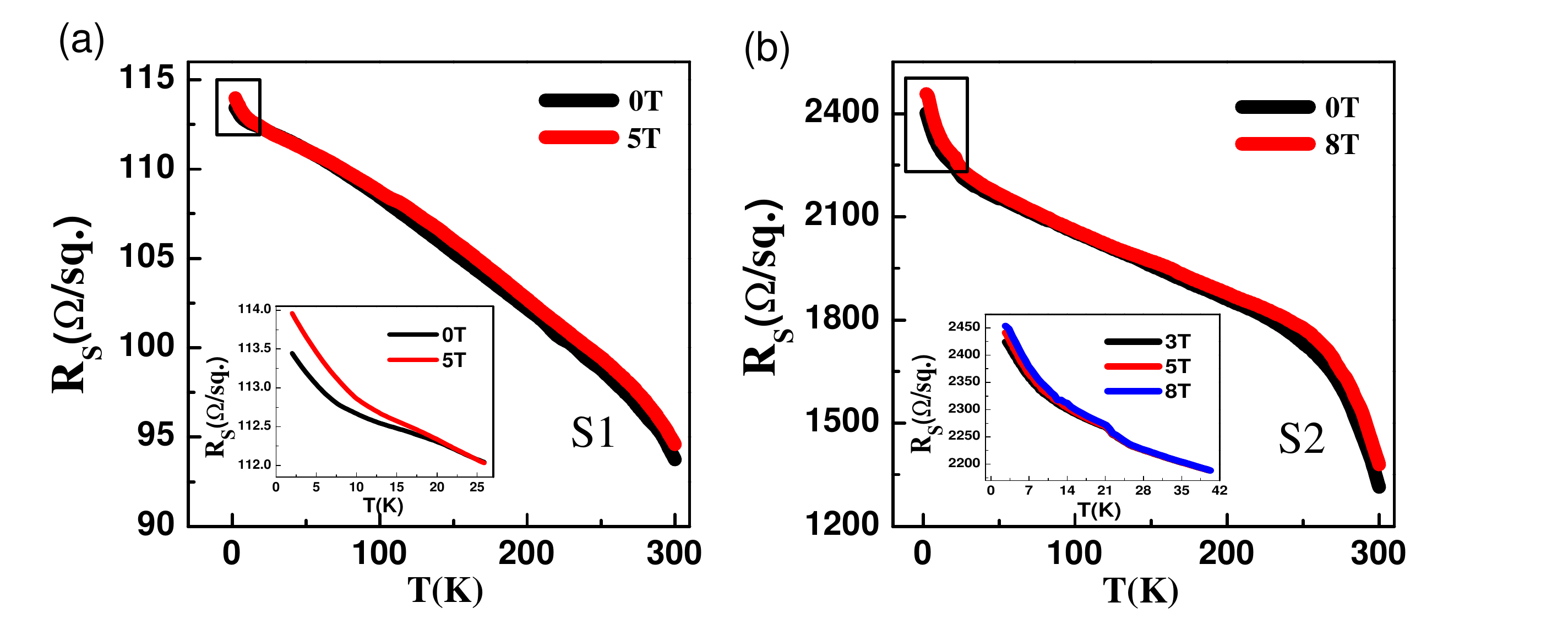}
	\caption{(a-b) Sheet resistance vs temperature at various magnetic fields for S1 and S2 films. Insets show an upturn in sheet resistance at low temperatures arising from electron-electron interactions.}
	\label{fig:rt}
\end{figure}

Figure~\ref{fig:rt} shows the variation of sheet resistance (R$_{S}$) with temperature for PS (S1, S2) thin films in various magnetic fields perpendicular to the film. We observe non-metallic behaviour for our PS thin films which is in contrast to the metallic behavior reported for the bulk materials.\cite{Chapai}  We speculate that this non-metallic behavior is a result of the quasi-low-dimensionality of the films.  This is supported by the fact that the resistance of the thinner film S2 (10~nm) is more than an order of magnitude larger than S1 (18~nm) as seen in Fig.~\ref{fig:rt}.  A change from metallic to insulating behaviour on reducing the thickness of various topological materials from bulk to quasi-two-dimensional has been reported previously.\cite{Kim2011,Liu2011, Zhao}  Below $20$~K, there is a pronounced upturn in the sheet resistance. This is highlighted in the inset of Fig.~\ref{fig:rt}. This enhanced upturn may either be due to electron-electron interaction in the two-dimensional limit or disorder-induced weak localization (WL).\cite{Hikami,JWang,Liu}  We will return to discuss this low temperature upturn in detail later.

\begin{figure}
	\centering
	\includegraphics[width=1.1\linewidth, height=0.22\textheight]{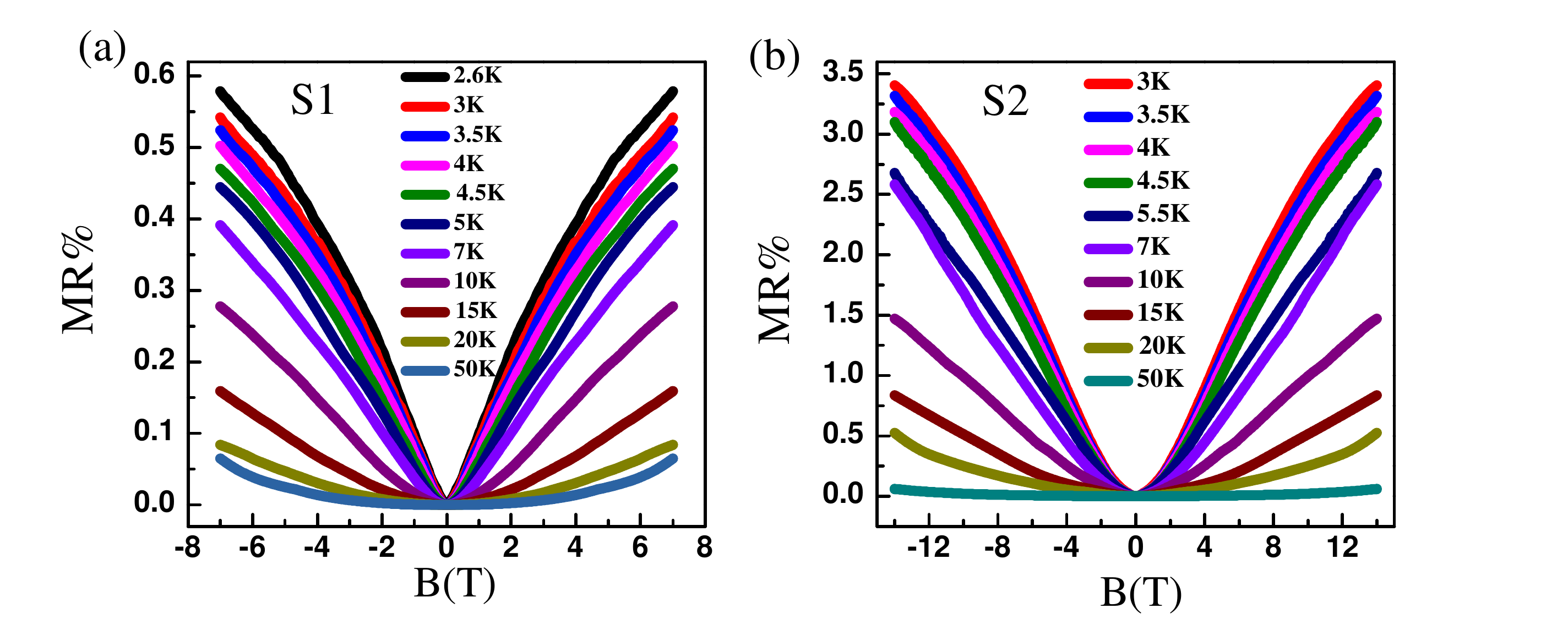}
	\caption{ The magneto-resistance (MR\%) vs out of plane magnetic field (B) at various temperatures for S1 and S2 films. }
	\label{fig:mr}
\end{figure}

\begin{figure*}
	\centering
	\includegraphics[width=1\linewidth, height=0.6\textheight]{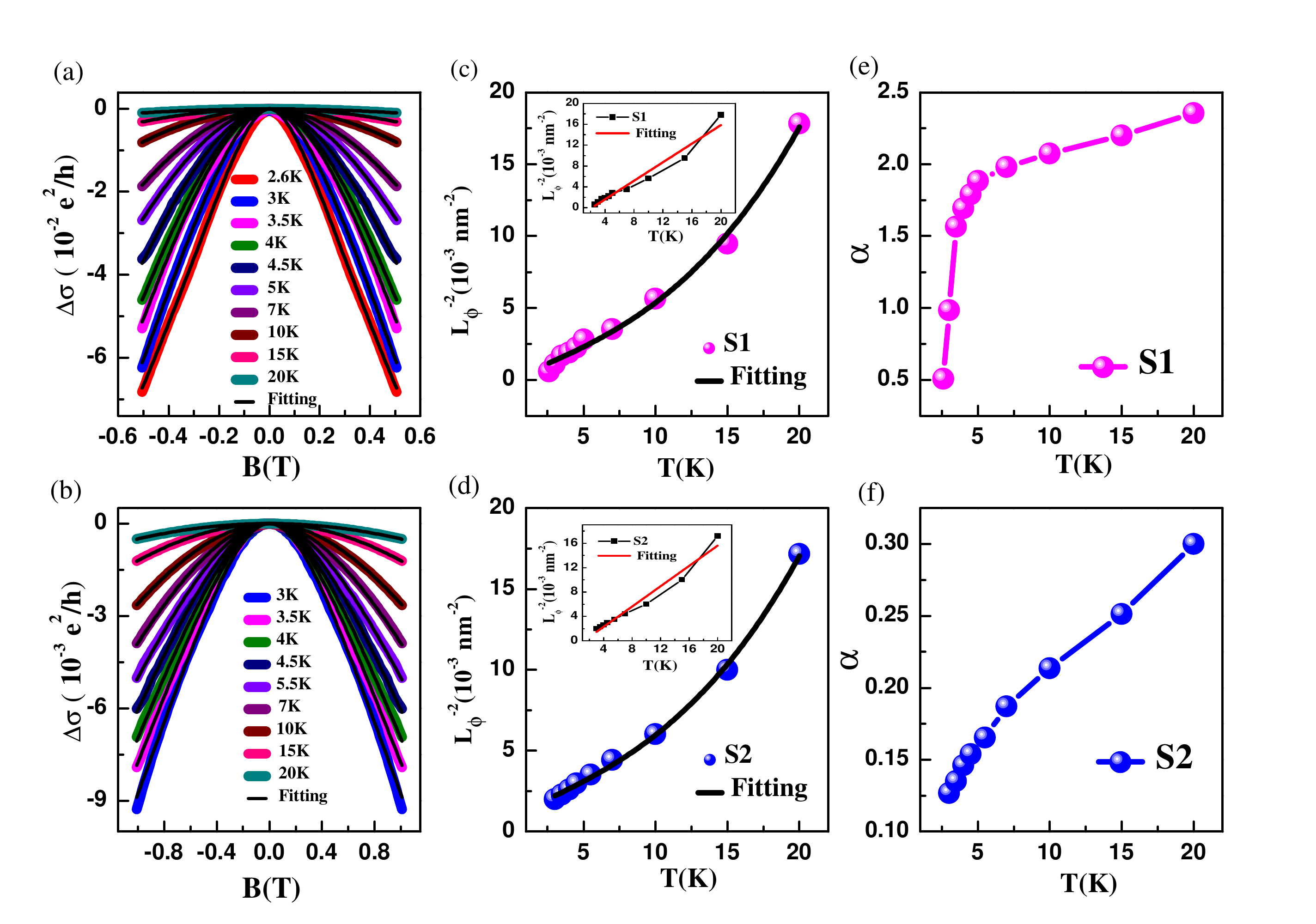}
	\caption{(a,b) Magnetic field dependence of magneto-conductance ($\triangle \sigma$) at various temperatures for S1 and S2. Black lines show fits with the HLN equation. (c,d) Variation of L$_{\phi}$ as a function of temperature, revealing the contribution of different scattering mechanisms. The inset shows deviation from a linear dependence.  (e,f) Temperature dependence of $\alpha$.}
	\label{fig:10002000-mr}
\end{figure*}

Figure~\ref{fig:mr} shows the MR\% at various temperatures for transverse magnetic field configuration in the range of $\pm$ 7 T and $\pm$ 14 T for S1, S2 thin films of PS, respectively. The Magneto-resistance percentage is defined to be 
\begin{equation}\label{key0}$$\centering
MR\% = $\dfrac{R(B) - R(0)}{R(0)}$ $\times$ 100
$$\end{equation}
where R(B) is the sheet resistance in a magnetic field (B). The sharp cusp type behavior around zero magnetic fields is observed in both samples, indicating the WAL effect in PS thin films.\cite{Lee,Bergmann,Hikami,Fang} The value of the magneto-resistance is low in comparison with single crystals,\cite{Chapai} which results from a much smaller carrier mobility in PS thin films as estimated from our Hall data discussed later. 

Before discussing the WAL, we should consider the dimensionality of the thin films. For Quantum interference (QI) effects, the relevant length scale is phase coherence length (L$_{\phi}$) which is equal to $\sqrt{D\tau}$, where $D$ is the diffusion constant, and $\tau$ is the phase coherence time. Also, the criterion for the 2D nature of thin films is L$\phi$ $>$ t, where 't' is the thickness of the film.\cite{Anderson,Ovadyahu} In our case, the value of L$_{\phi}$ is greater than the thickness of thin-film (based on the analysis below), indicating the 2D nature of thin films.

 \begin{figure*}
	\centering
	\includegraphics[width=1\linewidth, height=0.5\textheight]{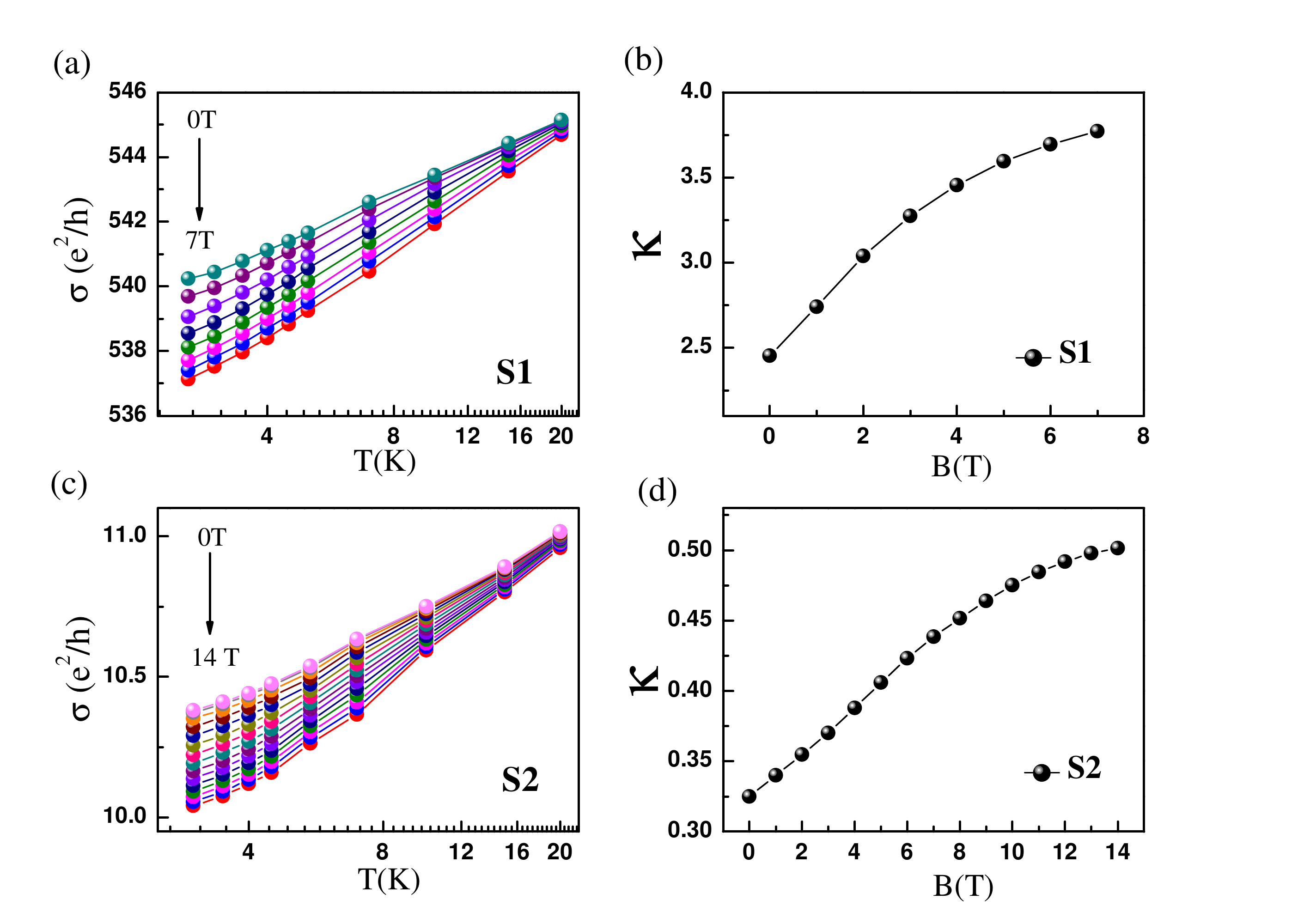}
	\caption{(a,c) shows the logarithmic temperature dependence of conductivity at low temperatures. The solid lines are guides for the eye. (b-d) shows the $\kappa$ versus $B$ obtained by linear fitting the conductivity data shown in (a,c).}
	\label{fig:eei-effect}
\end{figure*}

Figure~\ref{fig:10002000-mr}(a) shows the magnetic field dependence of conductance at various temperatures in the range of $\pm$ 0.5 T for the S1 sample. The two dimensional conductance was found using $\triangle$$\sigma$ = $\sigma$(B)-$\sigma$(0) where $\sigma$(B) = (L/W) (1/R$_{s}$), L and W are the length and width of sample respectively, R$_{s}$ is the sheet resistance. For two-dimensional systems, Hikami-Larkin-Nagaoka (HLN) equation can be used to describe the effect of localization.\cite{Hikami,Fang}. The HLN equation used for 2D systems is given by
\begin{equation}\label{key1}$$
\centering
$\triangle$ $\sigma$(B)= -$\alpha$$\dfrac{e^{2}}{\pi h}$\bigg[$\psi$\bigg($\dfrac{1}{2}$+$\dfrac{B_{\phi}}{B}$ \bigg)- ln \bigg($\dfrac{B_{\phi}}{B}$\bigg)\bigg]
$$\end{equation}
where $\psi$(x) is the digamma function, e is the electron charge, h is the Planck constant, B$_{\phi}$ is the characteristic field, which is required to destroy phase coherence. The characteristic field is associated with phase coherence length as B$_{\phi}$ = $\hbar^{2}$/(4eL$_{\phi}^{2}$). The parameter $\alpha$ is related to the number of conducting transport channels in a material. According to the HLN theory, parameter $\alpha$ takes value 1/2 and -1 for weak anti-localization (WAL) and localization (WL), respectively.

By fitting the experimental data to Eqn.~\ref{key1}, L$_{\phi}$ and $\alpha$ are extracted for S1 as shown in Fig.~\ref{fig:10002000-mr}(c,e). Figure~\ref{fig:10002000-mr}(c) shows the temperature dependence of L$_{\phi}$ for S1, which decreases from 41~nm at 2.6~K to 7.5~nm at 20~K\@. The Nyquist electron-electron theory predicts that L$_{\phi}$ $\propto$ T$^{-n/2}$ where $n = 1$ for 2D systems.\cite{Wu} Figure~\ref{fig:10002000-mr}(b) inset curve shows that L$_{\phi}$ deviates from a T$^{-1/2}$ dependence. This suggests that scattering mechanisms other than electron-electron scaterring are involved in dephasing the electron's phase in thin films of PS. To analyze the temperature dependence of L$_{\phi}$ we will fit using a simple equation:\cite{Bird}
\begin{equation}\label{key2}$$\centering
$\dfrac{1}{L_{\phi}^{2}}$ =$\dfrac{1}{L_{\phi o}^{2}}$ + A$_{ee}$ T$^{n}$ + B$_{ep}$ T$^{n'}$ 
$$\end{equation}
where $L_{\phi o}$ is the  temperature-independent dephasing length, A$_{ee}$ T$^{n}$ and B$_{ep}$ T$^{n'}$ represent the contributions from electron-electron (e-e) and electron-phonon (e-p) interactions, respectively. The value $n = 1$ is generally used for electron-electron scattering.\cite{Wu} According to electron-phonon interaction theory $n'\ge 2$.\cite{Sergeev,Reizer,Breznay}  Fitting the L$_{\phi}$ vs $T$ data by Eqn.~\ref{key2} with $n = 1$ we obtained an excellent fit with $n' = 3$ as illustrated by the solid black curve through the data in Figure~\ref{fig:10002000-mr}(c).  This reveals the presence of both e-e scattering and e-p scattering mechanism. The value of A$_{ee}$ = 4.12 $\times$ 10$^{-4}$ is very large as compared to the value of B$_{ep}$ = 1.2 $\times$ 10$^{-6}$, which suggests the dominance of the e-e scattering mechanism.  The value of $\alpha$ is $0.5$ at $T = 2.6$~K, which is in accordance with the theoretical value for a single conduction channel. Figure~\ref{fig:10002000-mr}(e) shows the variation of $\alpha$ as a function of temperature. It is clear from Fig.~\ref{fig:10002000-mr}(e), the number of conducting channels is strongly temperature-dependent.  This indicates that at low temperatures there is good coupling between the surface topological conduction channels and the bulk conduction channels, leading to a single channel.  As temperature is increased, a decrease in coupling between the various conduction channels occurs leading to the value of $\alpha$ to increase with increasing temperature.\cite{Wang2016, Sahu}  Above $\sim 10$~K $\alpha$ saturates to a value consistent with the number of conduction channels being 4-5.  

Qualitatively similar results were obtained for S2 as shown in Fig.~\ref{fig:10002000-mr}(b), (d), and (f).  Figure~\ref{fig:10002000-mr}(d) shows the temperature dependence of L$_{\phi}$ for S2, which decreases from 23 nm at 2.6~K to 7.6 nm at 20~K. The value of $\alpha$ is 0.13 at T = 3~K, which is less than the expected value. The low value of $\alpha$ most likely indicates the presence of both topologically trivial and non-trivial electrons,\cite{Shama} or enhanced disorder in the thinner S2 films as has been observed previously for disordered thin films of other topological materials. \cite{Brahlek2015, Liao2015}  

Now we turn to the low temperature upturn in the resistance which was highlighted in the insets of Fig.~\ref{fig:rt}.  The WAL dominated conductivity is expected to increase with lowering temperature without any external magnetic field.\cite{Shen} However, the measured conductivity drops logarithmically as temperature is lowered, as shown in Fig.~\ref{fig:eei-effect}(a). This could arise either from disorder-induced weak localization (WL),\cite{Hikami} or from electron-electron interaction effects.   For the WL effect, magneto-conductivity should be positive, which is not observed in our case. This suggests that electron-electron interaction could be the possible mechanism for the anomalous rise in the resistance at low temperature.\cite{JWang,Liu,Chen,Takagaki,Chiu} The  logarithmic T dependence of conductivity rules out the 3D theory for which we expect $\triangle \sigma = \sqrt{T}$.\cite{Lee,P}  Theoretically, 2D EEI correction to conductivity is given by
\begin{equation}\label{4}
\delta\sigma = - \dfrac{e^{2}}{\pi h}n\Bigg( 1- \dfrac{3}{4}F\Bigg)ln\Bigg(\dfrac{T}{T_{o}}\Bigg) = - \dfrac{e^{2}}{\pi h} \kappa ln\Bigg(\dfrac{T}{T_{o}}\Bigg) 
\end{equation}
where n is no. of conduction channels and F ($0 \leq $ F $\leq 1$) is the screening factor, and T$_{o}$ is the characteristic temperature for the EEI effect. 
Figure~\ref{fig:eei-effect}(a) shows the logarithmic temperature dependence of the conductivity for the S1 sample at various magnetic fields up to 7T. By applying Eq.~\ref{4}, $\kappa$ is obtained and is shown in Fig.~\ref{fig:eei-effect}(b) as a function of the magnetic field. On increasing the magnetic field, the slope of the $\sigma$(T) curve initially increases and then tends to saturate for higher fields. This indicates that at high fields, the WAL is quenched. Then the saturated $\kappa$ at high fields, henceforth called $\kappa_{ee}$, only includes the EEI correction to the conductivity.  For S1 then, $\kappa_{ee} \approx 3.76$ from Fig.~\ref{fig:eei-effect}(b).  A value of $F \sim 0.15$ has been reported for thin films of some topological materials like Bi$_2$Se$_3$. \cite{JWang} From the relation, $\kappa$ = n(1-3F/4), we obtain $n \sim 4$, corresponding to four transport channels.  This value is consistent with the estimates of $n = 4$--$5$ made from the WAL effect above.

Qualitatively similar results are obtained for S2 for which measurements were done in fields up to $14$~T\@. Specifically, quenching of the WAL effect for high fields was observed as shown in Figs.~\ref{fig:eei-effect}(c). By applying Eq.~\ref{4}, $\kappa$ is obtained and is shown as a function of field in Figs.~\ref{fig:eei-effect}(d). The high field value $\kappa_{ee}$ = n(1-3F/4)$\approx 0.5$ can be used to estimate the number of conduction channels.  We have seen previously from the WAL analysis that the value of $\alpha$ was smaller than expected for a single conduction channel.  This is most likely due to the enhanced disorder in the thinner film. Assuming $n = 1$, corresponding to a single transport channel, we obtain F = 0.66, which is much larger than the value for S1 estimated above but is within the acceptable range of $0 \leq F \leq 1$.  A high F$\sim 0.35$ have been previously reported for disordered WTe$_2$ films.\cite{Xurui}

\section{Conclusion}
In conclusion, PS thin films with different thicknesses were grown on Si (100) substrate by the pulsed laser deposition (PLD) technique. To the best of our knowledge, PS thin films have been grown for the first time. We find that the magneto-conductivity at low temperatures has contributions from both the WAL and the EEI effects. An analysis of the WAL effect gives an idea about the thickness and temperature dependent coupling between various conduction channels (Topological surface state and trivial bulk states) in the material.  Our results also elucidate the electron dephasing mechanisms.  We find that both electron-electron and electron-phonon scattering mechanisms were responsible for electron dephasing effects in PS thin films. The EEI effect is also observed at low
temperatures. From the EEI effect, we extracted the number of transport channels and found estimates in accordance with values obtained from the WAL effect. This suggests that the EEI effect can be used as an alternate method to estimate the number of conduction channels in materials where the WAL may not be observed.  

\section{acknowledgment}
We acknowledge use of the PPMS Sonipat (CFMS 14T, Cryogenic Ltd.), Central Research Facility, IIT Delhi for some transport measurements.


\end{document}